\documentclass[utf8]{frontierHLTH} 
\usepackage{url,hyperref,lineno,microtype,subcaption}
\usepackage[onehalfspacing]{setspace}

\def\keyFont{\fontsize{8}{11}\helveticabold }
\def\firstAuthorLast{Christoph Bandt} 
\def\Authors{Christoph Bandt} 
\def\Address{Institute of Mathematics and Computer Science, University of Greifswald, 17487 Greifswald, Germany} 
\def\corrAuthor{Christoph Bandt}
\def\corrEmail{bandt@uni-greifswald.de}

\begin{document}
\onecolumn
\firstpage{1}

\title[Turning rate and continuous hypnograms]{Crude EEG parameter provides sleep medicine with well-defined continuous hypnograms} 

\author[\firstAuthorLast ]{\Authors} 
\address{} \correspondance{} \extraAuth{}
\maketitle

\begin{abstract}
\section{}
To evaluate EEG data, one can count local maxima and minima on a fine scale, in a sliding window analysis. This straightforward calculation, which simplifies and improves previous work on permutation entropy, directly defines a good proxy for brain activity in an EEG channel during an epoch of 30 seconds. Different channels and persons can be compared when they are measured with the same device and prefiltering options.  This could lead to a rigorously defined and suitably standardized biomarker of cortex activity, like blood pressure or laboratory values. 

Applied to sleep EEG, the algorithm yields hypnograms with continuous scale which show amazing coincidence with sleep stage annotation by trained experts. Although produced by a crude method, continuous hypnograms provide a lot of details. For example, sleep depth usually decreases from evening to morning even within the same annotated sleep stage, except for REM phases where mean sleep depth is rather constant, but different in frontal and parietal channels. The diagnostic potential of the method is demonstrated with two hypnograms of narcoleptic patients. In all 10 subjects, infra-slow oscillations of activity with a wavelength between 30s and two minutes were clearly seen, particularly strong at the onset of sleep and in S2 phases.

The suggested method needs to be checked and improved. In its present form it seems already an appropriate tool for screening  long-term EEG data.

\tiny \keyFont{ \section{Keywords:} EEG, hypnogram, REM sleep, turning point, infra-slow waves, permutation entropy} 
\end{abstract}

\section*{Introduction}
EEG interpretation is an art that has developed over decades through careful observation of thousands of patients.  Its methodology is based on a variety of visual elements, and on the frequency spectrum with specific wave bands. In sleep medicine, spindles and $\delta$ waves are main features for classification of sleep stages.  Polysomnographic records, in particular electrooculography (EOG) and electromyography (EMG), are used for accurate detection of spindles as well as for distinction of REM phases. Sleep scoring requires well-trained, certified experts, and guidelines contain lots of rules \cite{RK,AASM07,AASM12}. Now there are many automatic sleep score programs which use a variety of complicated machine intelligence algorithms to imitate the expertise of sleep physicians \cite{Anderer05,Khalighi13,Lajnef15,Punjabi15,Lew}. Different specialists and computer packages usually agree on classification of easy cases, but often differ on complicated cases \cite{RV,Younes16,Pen}.

Some recent EEG studies use permutation entropy \cite{BP} in the analysis of sleep \cite{KL,NG,Ba17a}, epilepsy \cite{Bru,Fe}, Alzheimer's disease \cite{Mor}, and drug effects \cite{OSD}. The fine-scale background noise, usually removed by digital filtering, was shown to contain significant information which is not visible to the physician's eye.  
The present paper greatly simplifies and improves the methodology of these studies.  

In the method section we define {\it turning rate} as relative amount of local maximum and minimum points in an epoch.  
This unconventional, extremely simple calculation apparently defines a good proxy for brain activity in an EEG channel in a direct way.  For sleep medicine, the algorithm produces continuous hypnograms which are rigorously defined, without relying on any human or machine intelligence. Nevertheless, they contain more detail information than usual hypnograms. 
Turning rate should apply to EEG records in various other fields of medicine and psychology. 

This paper invites the reader to check and improve our method which is presented here in the simplest possible version. We argue by discussing features of several continuous hypnograms from the classical CAP database of Terzano et al. \cite{Te} available at physionet \cite{physio}. We are well aware that our findings remain mere hypotheses as long as they are not reproduced by others, with their own data. 

\begin{figure}[h!]
\begin{center}
\includegraphics[width=18cm]{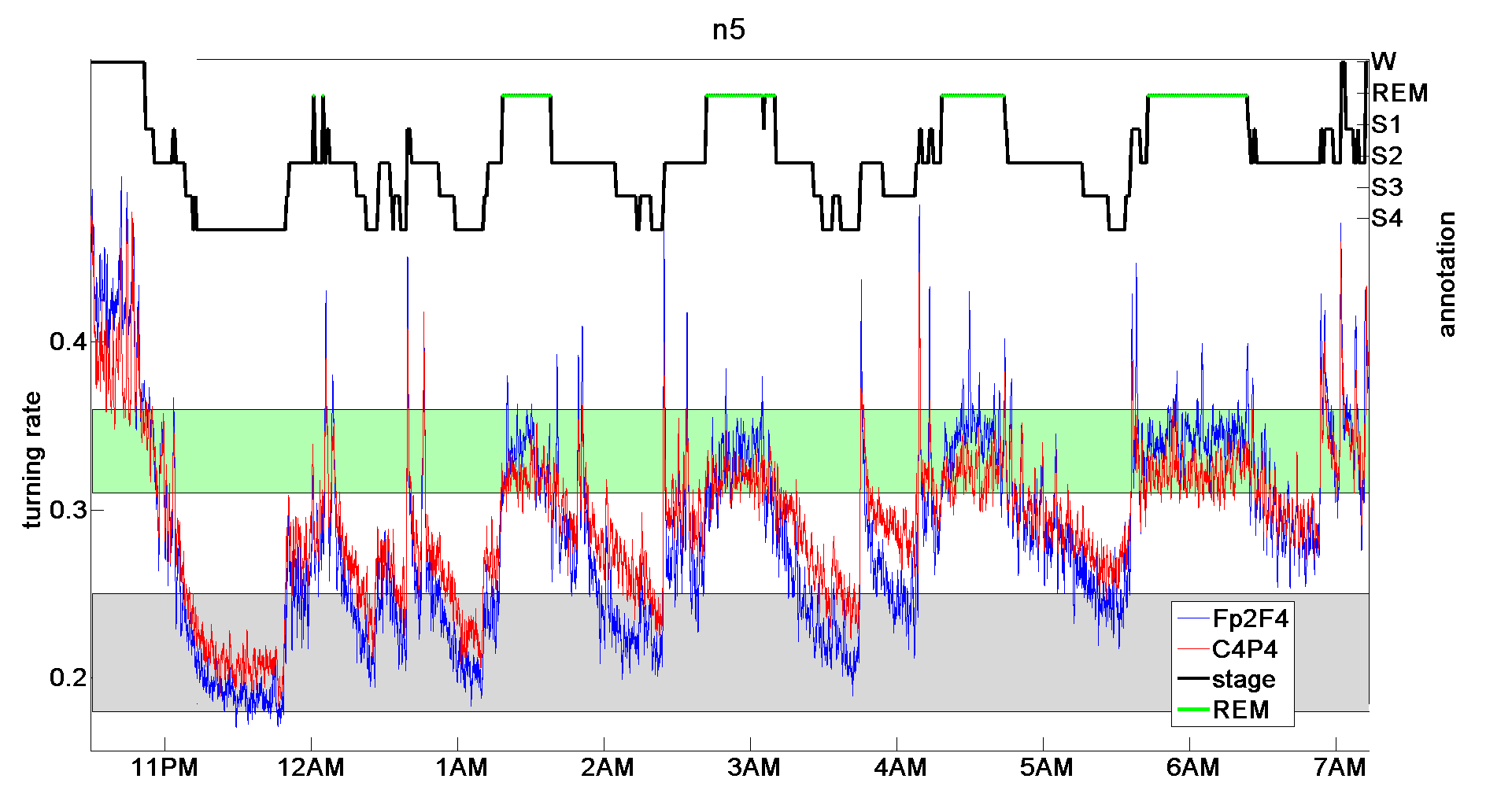}
\end{center}
\caption{Continuous hypnogram of the healthy control person n5 (lower part), compared to the annotation of sleep stages by a trained expert (upper part). Colored strips indicate range of turning rate in REM and SWS}\label{fig:1}
\end{figure}

Figure \ref{fig:1} shows a first example, based on data of a 35 years old healthy woman called n5 in the CAP database. The upper part shows sleep stage annotation by a trained expert, based on EEG potentials Fp2F4 and F4C4 and EOG channel ROC-LOC, using the R{\&}K rules which still included stage S4.   The lower part shows turning rates for the frontal channel Fp2F4 and the parietal channel C4P4 in our study.

The coincidence is amazing. Actually, we studied tens of patients and did not find a healthy case where coincidence at first glance was bad.  However, our intention is not to invent another automatic scorer! 
Turning rate and  continuous hypnograms are
rigorously defined, without a long list of rules and without any expert or machine intelligence. 
So turning rate could be a step towards an 
index of brain activity in an EEG channel similar to measurement of blood pressure.  

The name `continuous hypnogram' indicates that we measure activity, and hence sleep depth, on a continuous scale between 0 and 1 while sleep stages represent a discrete scale with five or six values only.  Asyali et al. \cite{Asyali07}, McKinney et al. \cite{McKinney11} and Younes et al. \cite{YounesOstrowski} also developed measurements of sleep depth on a continuous scale, based on experiments and analysis of frequency bands.  Our approach is much simpler, however, and comes in a natural way with the calculation.

As a rule in Figure \ref{fig:1}, turning rate above 0.36 means wakeful activity, the four REM  phases all have the same level between 0.31 and 0.36,  and values below 0.25 indicate slow wave sleep (SWS - that is, stage S3 and S4). Other persons show slightly different values. It seems that even a very active mind in the middle of the day will never reach values above 0.7. Already 0.6 indicates abnormal strong activity, or muscle artifacts . 
Our lowest observed values in sleep were around 0.15, for person n10 below. Still smaller values could occur for drug-induced sleep, anaesthesia, and koma. Turning rate 0 would theoretically correspond to brain death, but our method is not appropriate for extremely flat EEG.  

Compared to expert annotation, turning rate gives additional information in various ways. The following features of Figure \ref{fig:1} have been observed for most other patients in the study.

\begin{itemize}
\item Sleep depth on the continuous scale of turning rates leads to finer observations. For instance, turning rates with annotated S4 range from 0.18 to 0.27, with gradual increase from evening to morning.
\item The onset of sleep in the evening is fast, turning rate decreases like a parabola. After REM phases, the  turning rate decreases only linearly towards slow wave sleep, with smaller slope in the morning.
\item In slow wave sleep, the frontal channel is less active than the parietal one. In REM sleep and wakefulness, however, the frontal channel becomes more active. Actually, for most patients REM phases can be defined by this property.
\end{itemize}

The last remark indicates that turning rate may be used as a tool for comparing different channels. This could lead to applications in areas where different positions have to be compared or a spatial focus exists: epilepsy, stroke, injuries, recovery after brain operation. In the present paper, however, we only consider the above two channels in sleep EEGs.

Continuous hypnograms show lots of fluctuation, partly caused by statistical error and artifacts , but mostly due to interesting detail structure. Close-ups of Figure \ref{fig:1} will be studied in the results section. In particular, we shall see that there are very interesting infra-slow waves of activity with a wavelength between 30 seconds and 2 minutes. They seem particularly strong at the onset of sleep and at S2 phases after slow wave sleep, for instance just before midnight in Figure \ref{fig:1}. These waves deserve further study.  

The method seems promising.
Of course, it has to be checked with other data, more EEG channels, repeated measurements of the same subject etc. It would be nice to understand how turning rate depends on prefiltering options of the record, in order to perform a transparent calibration of data measured under different conditions. It would also be desirable to define brain activity in epochs much shorter than 30 seconds.  Actually, we shall calculate turning rates for epochs of 1s but their statistical error forces us to do some smoothing over 30s. Other ideas are required to measure change of brain activity within parts of a second, which would be needed for application to evoked and event-related potentials.

\section*{Method}

\subsection*{Introduction to turning rate} 
We consider a time series of values $x_1,x_2,...,x_T.$ A time point $t$ is called {\it turning point} if there is 
\[\mbox{either a strict local minimum, } x_{t-1}>x_t<x_{t+1}\, , \mbox{ or a strict local maximum, } x_{t-1}<x_t>x_{t+1}\, .\]
The relative number of turning points in the series is called {\it turning rate.} 
\begin{equation}\label{tura}  \mbox{\it turning rate} = \mbox{number of turning points between 2 and $T-1,$ \ divided by } T-2 \ ,
\end{equation}
since we cannot decide whether time points 1 and $T$ are turning points.
Thus the turning rate is always between 0 and 1. It can be considered as a probability, or as a percentage between 0 and 100 \% . It has no measuring unit. However, it depends on the sampling frequency. In a high-resolution EEG, there are long increasing and decreasing parts, thus a small turning rate.  For low resolution, like 50 Hz, there will be more up and down, and a higher turning rate. Roughly speaking, we shall take the turning rate defined in \eqref{tura} at sampling frequency 128 Hz.

\begin{figure}[h!]
\begin{center}
\includegraphics[width=18cm]{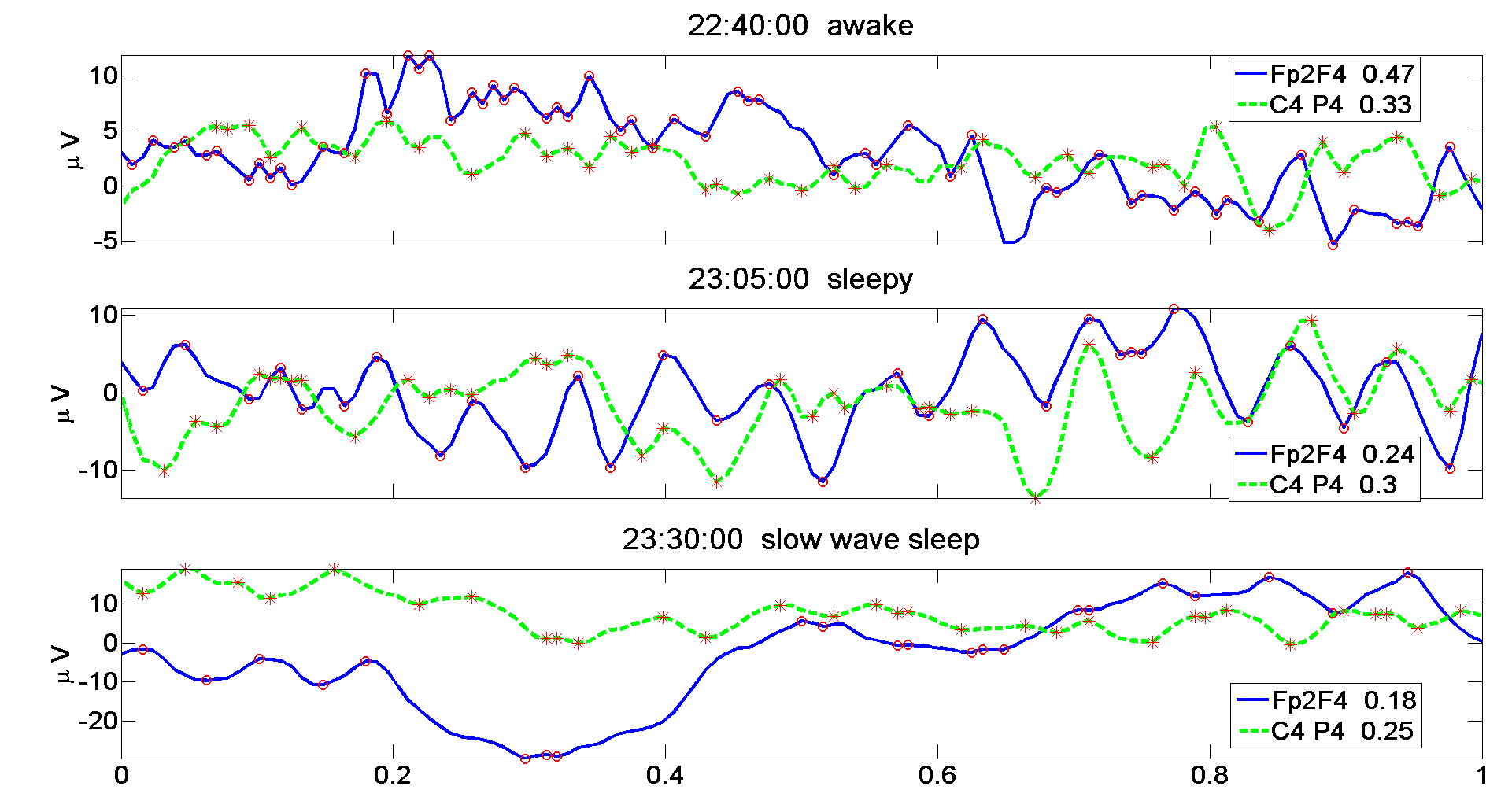} 
\end{center}
\caption{EEG samples of 1 second length, downsampled to 128 Hz, of the subject of Figure \ref{fig:1} in wakeful, sleepy, and slow wave state. Turning points are marked, turning rate is given in the legend.}\label{fig:2}
\end{figure}

Figure \ref{fig:2} shows how this works out for our healthy volunteer. If she is awake, the EEG shows rather small oscillations without a characteristic frequency, and more than every third time point is a turning point.  If she gets sleepy,  $\alpha$ waves and $\sigma$ waves (spindles) will start to dominate the small fluctuations, and the turning rate decreases. Finally, in slow wave sleep, we have big $\delta$ waves which suppress the tiny up-and-down, and the turning rate becomes still smaller, around 20 \% . 

Actually, even in the absence of $\sigma$ and $\delta$ waves, as for C4P4 in the lower panel, the turning rate reflects sleep depth. Here is a rough explanation.  All EEG potentials represent a complicated mixture of a lot of tiny electrochemical processes in the cortex, most of them of a duration of few milliseconds \cite[chapter 1]{ZH}. We do not exactly know what we measure. However, we can expect that some of the contributions cause turning points in the series of measurements, and we shall have more turning points when we have more contributions. In this way, turning rate of an EEG channel becomes a proxy of the electrochemical activity of the respective region of the cortex. Only experiments can verify to which extent this vague idea is correct.

\subsection*{Epoching and variance problem} 
The big time series covers several hours. It is divided into disjoint {\it epochs} of short length, and the turning rate is determined for each epoch. A length of 30 seconds corresponds to customary scoring, and was used in \cite{Ba17a} for a kind of permutation entropy. Here we use an epoch length of one second only, in order to analyze details of the continuous hypnogram. 

There is a problem with such short epochs which is well known for the frequency spectrum. It is the {\it variance} of turning rates which creates an unavoidable statistical error. Equation \eqref{tura} estimates the probability of the event ``turning point'' using a series of $T$ experiments, as probabilists say. From toin cossing it is known that the statistical error is about $1/\sqrt{T}$ (when the true probability is $\frac12 ,$ this is two times the standard deviation of the estimate, commonly used as radius of the 95\% confidence interval). This is just a rule of thump since our `experiments' are usually correlated. It is confirmed by 
figures in the results section which illustrate statistical error in our data.
For 128 Hz sampling frequency and one second epochs, the error is about 0.1 which would not allow us to distinguish sleep stages.  Thus a main task will be {\it variance reduction} of turning rate estimates which is done in various ways explained below.

\begin{itemize}
\item We use 512 Hz sampling frequency and compare time $t$ with $t-4$ and $t+4,$
\item we smooth the series of turning rates by a moving average of length about 30,
\item we take windows of variable time length which begin and end with zero-crossing of the data. 
\end{itemize}

Before we go on, we note that the statistical use of turning points is not new. The French statistician Bienaym\'e, born 1796, developed a turning point test in 1874 to decide whether a sequence of measurements significantly differs from independent random numbers (white noise), and proved that some measurements of his fellow astronomers failed the test \cite{Bien74,Bien75}. According to his work, the mean turning rate of white noise is $\frac23 .$ Turning rate of EEG series with sampling frequency larger 100 Hz is definitely much smaller, even in the wakeful state, and has not been considered so far.

\subsection*{Choice of delay}
A little problem was not yet mentioned: in \eqref{tura}, we have to exclude neighbor points with equal values. Let $T'$ be the number of time points $t$ between 2 and $T-1$ for which $x_t\not= x_{t-1}$ and $x_t\not= x_{t+1}.$ Then we have to divide by $T'$ instead of $T-2$ in  \eqref{tura}. In the upper panel of Figure \ref{fig:2}, for example, two neighboring minima are neglected since they are equal. As an alternative, one can add a tiny white noise to the signal which will destroy all possible equalities. For both options, the existence of many equal values will increase the variance of turning rate estimates, and thus decrease accuracy, by almost the same amount.

Since our data are sampled with 512 Hz, we make use of the large dataset to reduce variance. For some number $d=1,2,...$ we say that $t$ is a {\it turning point with respect to delay $d$}  if either $x_{t-d}>x_t<x_{t+d}$ or $x_{t-d}<x_t>x_{t+d}.$ Taking a certain $d$ is almost the same as sampling with frequency $512/d$ Hz, with the advantage that we still have the same number of almost $T$ time points for the estimate of turning rate.

\begin{figure}[h!]
\begin{center}
\hspace*{-5ex} 
\includegraphics[width=18cm]{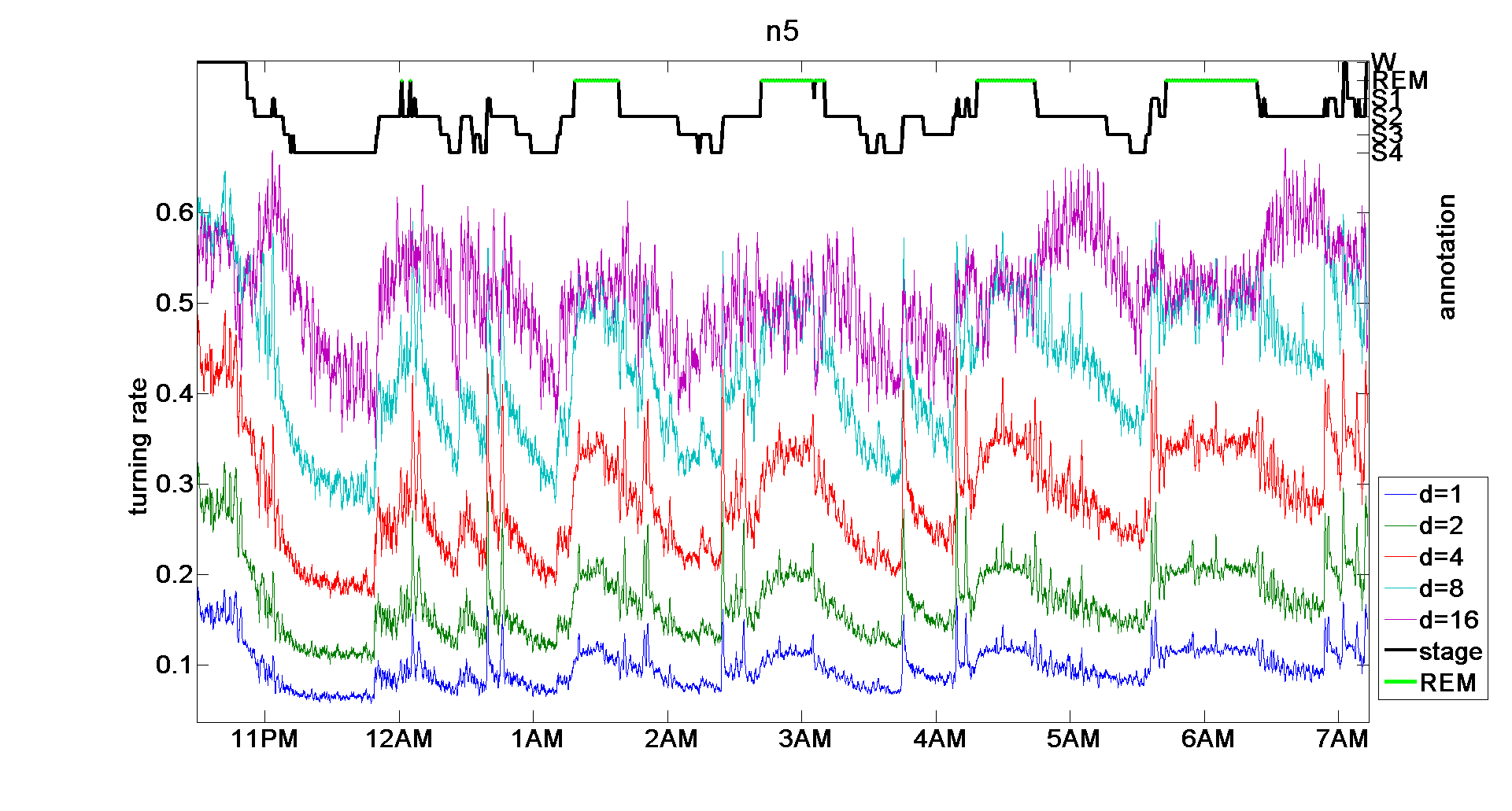} 
\end{center}
\caption{Continuous hypnograms of channel Fp2F4 of the subject of Figure \ref{fig:1} for different delays $d=1,2,4,8,16,$  corresponding to sampling frequencies 512, 256, 128,  64, and 32 Hz. We choose $d=4.$}\label{fig:3}
\end{figure}

One can ask which $d$ will work best. Figure \ref{fig:3} shows the continuous hypnogram of Figure \ref{fig:1} for $d=1,2,4,8,16.$ The level of turning rates is smaller for small $d,$ because of the smoothness of the series caused by the low pass prefilter. But the shape seems always the same, as long as we choose $d$ between 1 and 8. One should not take $d>10$ for 512 Hz sampling frequency because then turning rates are strongly influenced by the presence of $\sigma$ and $\alpha$ waves which cause the big variance for $d=16$ in Figure \ref{fig:3}. Turning rates for such $d$ can exceed the level $\frac23$ of white noise.  On the other hand, $d=1$ seems to be susceptible to small perturbations, since differences among neighboring values are so small. For $d=1$ we also had more than 6\%  of values with equal neighbors, which reduces to 1.7\%  for $d=4.$  

After testing some datasets, \emph{we decided to take $d=4$} which roughly corresponds to 128 Hz sampling frequency while utilizing the larger number of values to obtain higher accuracy. Other choices are possible.

\subsection*{Smoothing and artifacts} 
Seen from a mathematical viewpoint, our method transforms a huge series of about 15 millions of EEG measurements into a still fairly long series of 30000 turning rates of 1s epochs. This transformation is highly non-linear.  For this reason, it makes no sense to do preprocessing, like moving average or other filters, before determining turning rates, because we really do not know how the non-linear transformation  will modify the intended improvements. It makes sense, however, to {\it apply filtering after the transformation} since the turning rates of all epochs form an ordinary time series. 

We have chosen to smooth the series of turning rates by a \emph{moving average of length 31.} In effect this is the same as overlapping 30s epochs which are shifted by 1s only. The result is a smoothed series of several thousand values of turning rates, second by second. This will be used to study details of the continuous hypnogram, in particular to discover infra-slow oscillations with a wavelength between one and two minutes. On the other hand, our moving average suppresses oscillations with wavelength 30s or smaller which seem also to exist in the data.

{\it Artifacts} are a common problem in EEG analysis. Our method is influenced by two types of artifacts. Low frequency perturbations with sufficiently high amplitude remove turning points, as explained for $\delta$ waves in Figure \ref{fig:2}, and thus falsely indicate smaller activity. These artifacts were not so serious. Then there are muscle artifacts with rapid oscillations, greater than 30 Hz, often combined with changes of high amplitude, of duration 1s  up to 20s and more. Such artifacts cause a lot of turning points and can falsely pretend stronger brain activity. Some are very apparent as peaks in Figure \ref{fig:1}.  However, the worst of these artifacts  (when somebody turns around or moves the jaws for more than 10s) are actually connected with an annotated change of sleep state, with increase of activity. In Figure \ref{fig:1} they account for half of the peaks. The other half could be arousals, brain's reaction to a disturbance from inside or outside the body \cite{Parrino06}. As this has not been checked, it would be fair to mark all these places as gross artefacts instead of pretending accurate estimation of large turning rates.

There are many procedures for the detection and elimination of artifacts . For polysomnographic records, muscle and eye artifacts  can be controlled by the respective EMG and EOG channels. Intricate methods can remove artifacts directly in a multichannel EEG, see for instance \cite{CB00,Fitzgibbon07,Nolan10,Mognon11}.  A classical paper on eye artifacts in event-related potentials \cite{Gratton83} has been cited 3000 times which shows that this is a wide field.

{\it We did not include any artifact treatment} to keep our presentation simple.  Note that, since we have a  moving average of 31 one second epochs, artifacts of few seconds length would need only be detected and corresponding epochs treated as missing values, since neighboring epochs can fill the gap.

\section*{Data}
\subsection*{Data quality}
The method applies to standard raw EEG data.  Any dataset can be used for a rough trial.
For better accuracy, however, a few points should be observed, since we study the fine structure or 'background noise' of the signal. We found that a \emph{sampling frequency} of 128 Hz is sufficient to get all-night hypnograms like Figure \ref{fig:1}, but higher resolution is needed to study details. Our data were all sampled at 512 Hz. Moreover, it is important to note that the fine structure depends on filtering. We take raw data with \emph{minimal preprocessing} since excessive filtering can suppress the high-frequency structure. EEG data usually come with \emph{prefiltering} of the EEG device. Our data were measured with \emph{upper cutoff frequency 30 Hz and lower cutoff frequency 0.5 Hz.} Such a decent bandpass prefilter is needed for our method. On the one hand, data should be sufficiently smooth on the highest possible resolution, as seen in Figure \ref{fig:2}, since the turning rate for $d=1$ should not be too large. For our data, we could take larger upper cutoff, and compensate by choosing $d$ below 4 to obtain similar values. On the other hand, we should not have a strong baseline drift in the data which would decrease the turning rate. This is guaranteed by the lower cutoff.

There are two other important qualities of the data. The dynamic range of the record should be fully exploited so that many different values are possible and only \emph{few neighboring values are equal.} Such values are lost for the count of turning points, as noted above  Today's measuring equipment comes with at least 16 bit accuracy and this condition does not create a problem. We used old data with only 12 bit accuracy. Finally, there should be \emph{no 'mains hum' artifacts } from the power net since this would disturb structure in the high-frequency range which we consider. Other artifacts  will be taken as they are.

\subsection*{The database}
The data for this investigation were taken from the CAP sleep database of Terzano et al. \cite{Te} available at physionet.org \cite{physio}. The original purpose of Terzano et al. was the study of cyclic alternating patterns (CAP). The database contains 108 polysomnographic records with quite different data structure. The authors of the CAP study did experiment with the measurements, and one can feel their research spirit and passion for the subject. It was the pioneering time when the physionet databank started operation, and this author is very grateful to Terzano's group and to physionet for the well-documented data.

For most datasets in the CAP database, bipolar EEG montages were taken on the line Fp2-F4-C4-P4-O2 on the right side of the head, sometimes also at corresponding positions on the left. We agree with this choice since differences of nearby electrodes show little disturbance from distant sources, like power net, heart, and neck muscles. 
Some of the records contained up to 13 EEG channels, and many of them can be used to establish continuous hypnograms. Neighboring electrodes give better results than potential differences with large distance, like C4O1. 

For the present study, we decided to focus on a {\it very simple setting} with only two channels: the apparently most informative channel Fp2F4, and the more parietal channel C4P4 for comparison. We shall refer to Fp2F4 as `the frontal channel' and to C4P4 as `the parietal channel'.

We have chosen only records which contain EEG channels Fp2F4 and C4P4 with sampling frequency 512 Hz and at least six hours duration. 
Six of the 16 healthy controls satisfy this condition. Their continuous hypnograms will all be shown here, to give an impression of the variety of hypnograms for healthy sleep. Among patients with different symptoms, we restricted ourselves to four cases which demonstrate the diagnostic potential of continuous hypnograms.  

All chosen records fulfil the above quality criteria. They have the same prefiltering bounds 30 Hz and 0.5 Hz, which makes them comparable.
All selected datasets were professionally scored with special regard to EEG channels Fp2F4 and F4C4 where all CAPs had to be detected. The old R{\&}K rules with distinction between stages S3 and S4 stages were used for scoring. This was an advantage for us since in our continuous hypnograms we can distinguish various stages of sleep depth within slow wave sleep.

\begin{figure}[h!]
\begin{center}
\includegraphics[width=18cm]{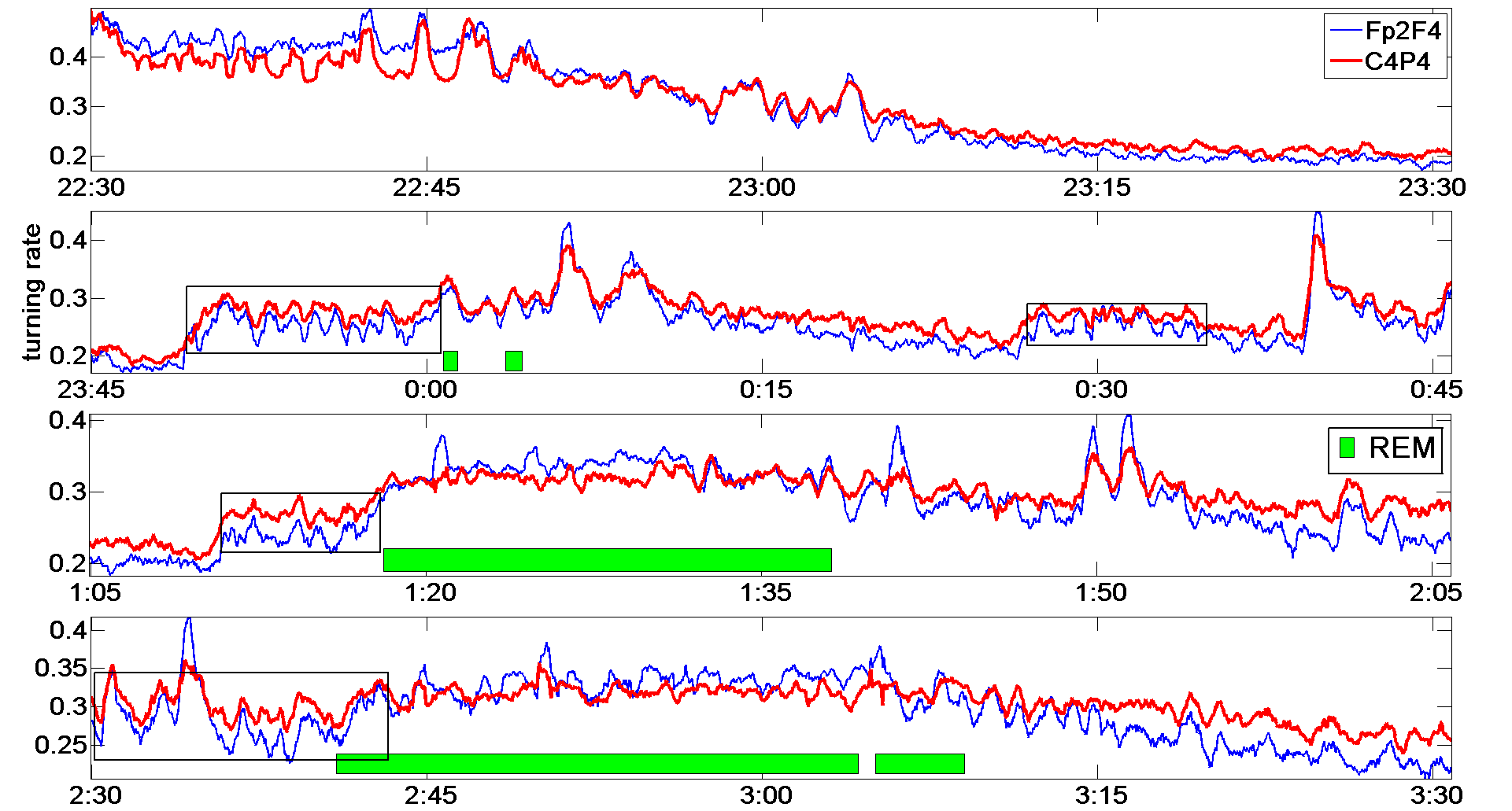} 
\end{center}
\caption{Four pieces of one hour length of the continuous hypnogram of n5. Infra-slow waves with wavelength between 30 s and 2 min are apparent at the onset of sleep in {\bf a}. They also appear within S2 blocks in {\bf b,c,d} but not within REM phases in {\bf c,d}.}\label{fig:4}
\end{figure}

\section*{Results}
\subsection*{Details of a continuous hypnogram}
In the following, the term `activity' is used as a synonym for `turning rate'. 
First we study the continuous hypnogram  of our healthy person n5 in greater detail. Figure \ref{fig:4} shows four pieces of one hour length. In  {\bf a} we see the transition from wakeful activity with turning rates above 0.4 to S4 sleep with turning rates below 0.2. Most obvious are two sets of waves around 22:45 and 23:00 pm. 

Oscillations of the turning rate with wavelength between 30 seconds and 2 minutes will be called $\tau $ {\it waves} for short. They appear in all subsequent figures, in particular at the onset of sleep and in S2 phases.  Such waves are known from electrophysiological recordings of animal cell populations  \cite{Aladjalova57,Hughes11}, from fMRI studies  \cite{Zuo10,Picchioni11,Mitra15}, intracranial records (ECoG), and local field potentials \cite{He08,Nir08}. In ordinary EEG they appear as periodic occurence of graphic patterns like spindles and CAPs \cite{Te,Parrino06}, and
as modulation of the power of traditional frequency bands \cite{Mantini07}. The wavelength between 30 and 120s is called slow-5 band \cite{Buzsaki04,Zuo10}. Most similar to our waves are the full-band EEG recordings presented by Vantahalo et al.  \cite{Vanhatalo04} and Monto et al. \cite{Monto08}, obtained with sophisticated equipment. See the discussion. 

In our standard EEG data, such oscillations come up regularly, similar to $\alpha$ and $\delta$ waves, only on the scale of turning rates. They usually appear in different channels at the same time, which indicates that they come from a subcortical region. One can speculate that they moderate the transition to deep sleep. When slow wave sleep is reached in Figure \ref{fig:4}{\bf a} after 22:10 pm, there are smaller and less synchronized oscillations.

Several blocks with sleep stage S2 show strongly synchronized $\tau $ waves with constant average. They are indicated by boxes in Figure \ref{fig:4}, first in {\bf b} around 23:50 pm after fast transition from S4 to S2, and then in {\bf c} after 1:10 am.  Both blocks end with increase of turning rates, and an `attempt' to get into REM sleep. In {\bf b} just after midnight there are two epochs  annotated as REM with the help of the EOG channel, but further $\tau $ waves provide the way to deeper sleep. This repeats with the S2 block around 0:30 am. In {\bf c} before 1:20 we finally get into a long REM phase with desynchronized channels.  Another constant S2 block with subsequent REM phase is shown in {\bf d}. 
Both REM phases are characterized by greater activity of the frontal channel, and both are followed by strong synchronous $\tau $ waves which lead back to deep sleep, with smaller activity of the frontal channel. 
Similar observations can be made for other probands. Many, but not all REM phases are preceded by an S2 block of constant mean activity.  

\begin{figure}[h!]
\begin{center}
\includegraphics[width=18cm]{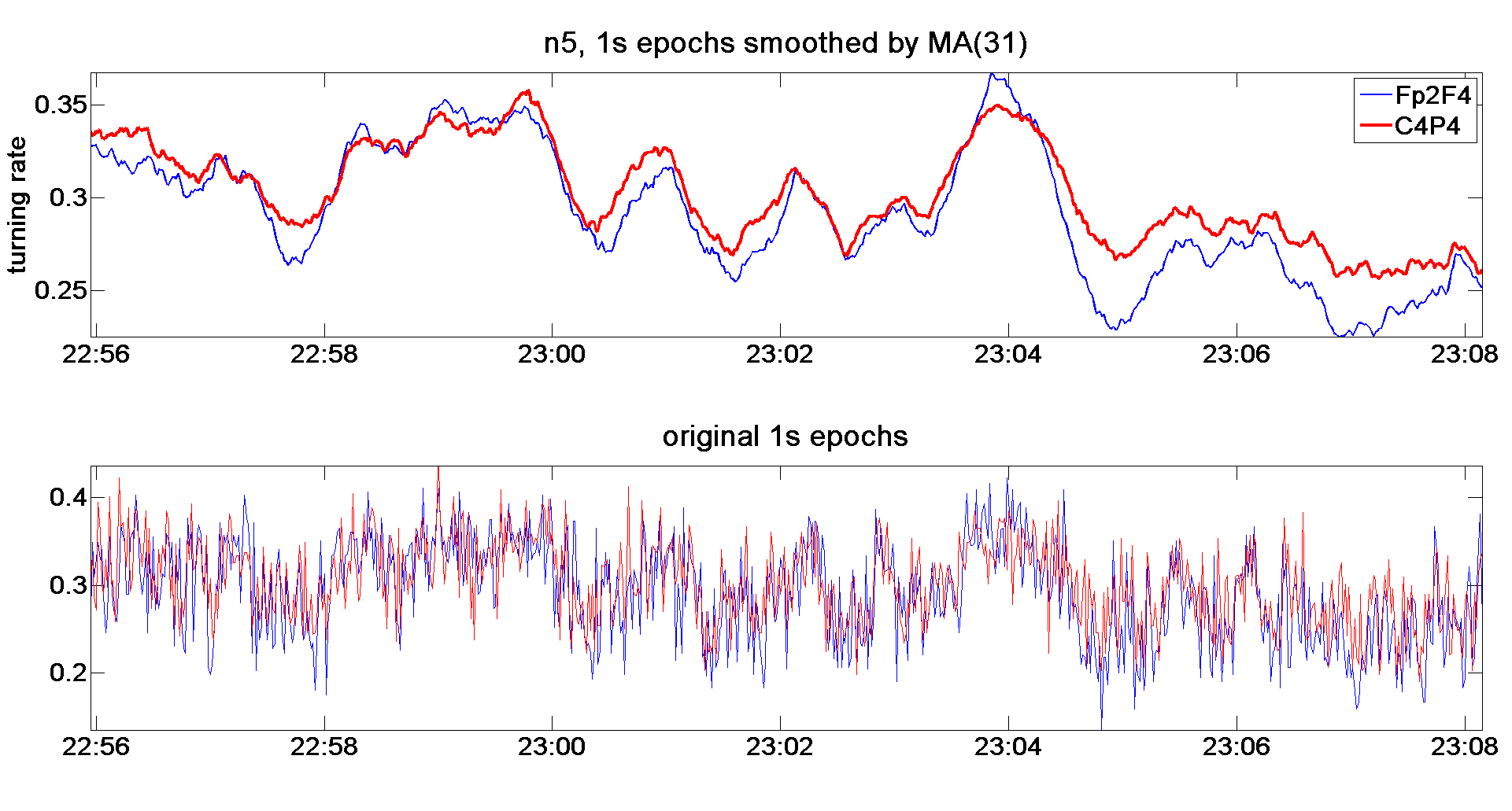} 
\end{center}
\caption{12 minutes from the upper panel of Figure \ref{fig:4}. Does our MA(31) smoothing make infra-slow waves visible or does it generate them?  Within this time window, infra-slow waves can be seen even in the original 1s epochs. The lower panel also shows the level of variance of these original turning rates. }\label{fig:5}
\end{figure}

Note that $\tau $ waves are oscillations of the turning rates and have no direct connection with the original data. Their wavelength is far outside the scope of usual EEG analysis. 
The frequency of an $\tau $ wave is between 0.03 and 0.01 Hz, and the high pass threshold of our data was 0.5 Hz.
The $\tau $ waves come with our turning rate methodology. Moreover, some of the biggest `waves' in panels {\bf b,c,d} of Figure \ref{fig:4} are connected with artifacts. Thus we must critically ask whether $\tau $ waves are perhaps an artificial product of our method, in particular of the moving average smoothing of 1s epochs. To this end, we investigate the waves around 23:00 in panel {\bf a} in greater detail in Figure \ref{fig:5}.

 In the lower panel, original turning rates of 1s epochs are shown. There are fast fluctuations, not synchronized in the two channels, which represent the variance of the statistical estimates of the turning rate. We also see slow oscillations with a  bigger amplitude which concern both channels together. In this figure, the $\tau $ waves become obvious for the original series of 1s epochs!  The upper panel shows 1s epochs smoothed by moving average of length 31. For each second $t,$ the average of all epochs between $t-15$ and $t+15$ is drawn, with the hope of getting the time series smooth and the $\tau $ waves more apparent. In fact, the erratic fluctuations of the lower panel disappear, and waves with wavelength 1 up to 2 minutes become visible. Not all of these long waves do exist in the lower panel. At 23:06, for instance, there seem to be two waves of length about 40s which in the upper panel become one wave of about 1.5 minutes.  Before 23:02 there is a similar pattern.  On the whole, however, smoothing removes random fluctuations and makes those long waves apparent which do already exist in the lower panel.

Our main argument for correctness of the smoothing procedure is the synchronization of the two channels in the upper panel of Figure \ref{fig:5}. The two EEG channels can be considered as independent measurements on different parts of the head.  When the resulting smoothed series of turning rates do almost coincide, the only plausible explanation seems that there must be a common source for the $\tau $ waves in both channels.  Throughout Figure \ref{fig:4}, with exception of REM phases and slow wave sleep, we have a similar synchronization.  The largest $\tau $ waves can be seen at the onset of sleep and at the beginning of blocks which lead to slow wave sleep. Smaller $\tau $ waves with very strong synchronization are within the S2 blocks marked by rectangles.

\begin{figure}[h!]
\begin{center}
\includegraphics[width=18cm]{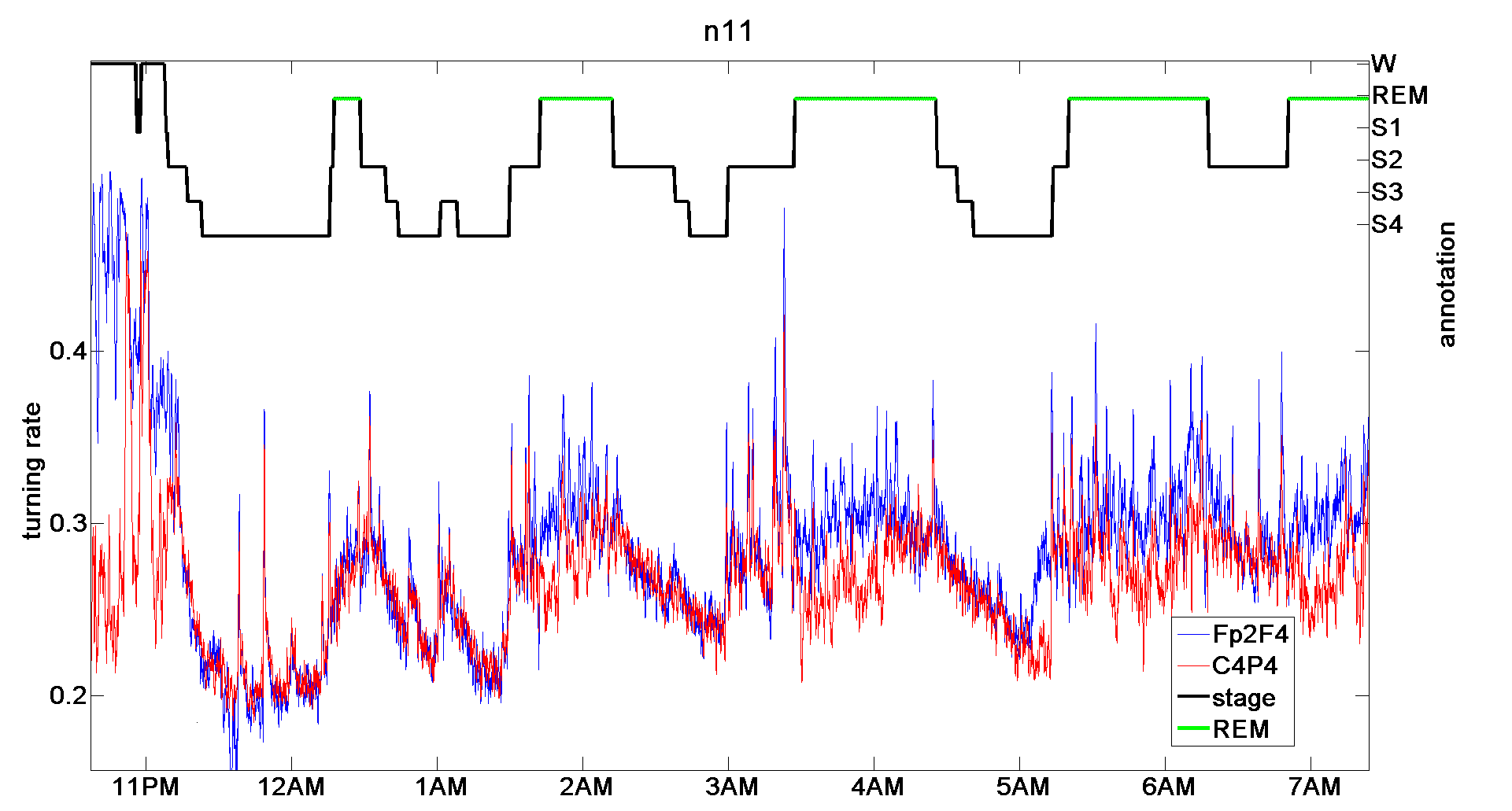} 
\end{center}
\caption{Continuous hypnogram of n11, with low activity in the parietal channel.}\label{fig:6}
\end{figure}

\subsection*{Main features of a continuous hypnogram}
Conventional hypnograms are based on observations of graphical elements in sleep EEGs ($\alpha$ waves, spindles, K complexes, $\delta$ waves). The main advantage of such hypnograms is their simple structure and schematic view.
What observations can be made with our method? Can continuous hypnograms also be simplified, in a scheme which avoids those nasty fluctuations? A first step to this end is done here. We give a list of properties, or better: hypotheses, derived empirically just by studying continuous hypnograms, without exploiting knowledge of sleep annotation. Seen from this viewpoint, the blocks of a continuous hypnogram differ a bit from conventional sleep stages. In the following, we disregard fluctuations and concentrate on essential trends of activity in a time window of at least 5 minutes, say.

\begin{enumerate}
\item  A continuous hypnogram consists of two types of building blocks, those with decreasing activity and those with constant activity. Transitions between blocks are fast and will be considered as sudden jumps or interruptions.
\item  A decreasing block always leads to deeper sleep. In conventional notation, it is part of a sequence S1-S2-S3-S4, depending on when it  begins and ends.
\item  There are three types of constant blocks: wakeful activity, REM, and constant S2 blocks.
\end{enumerate}

When a proband is awake, brain activity is of course not constant. For our measurements, however, where probands are in bed, this seems to be a reasonable simplification, as can be seen in the figures. Likewise, REM activity is not completely constant. In Figure \ref{fig:6} activity increases towards the end of REM phases. In Figure \ref{fig:1} and the figures below, however, constant REM level is a reasonable simplification. The need to consider transitions as jumps comes from the statistical nature of our method: we are not able to investigate fast changes. When the subject is awake, jumps of activity can lead up and down. In sleep phases, however, jumps will only lead upwards:

\begin{enumerate} \setcounter{enumi}{3}
\item  The decrease of activity in sleep is always gradual and continuous (and on closer look modulated by $\tau $ waves) while the increase is fast and often instantaneous.
\item  The fastest decrease of activity appears at the onset of sleep. The first decreasing blocks have a parabolic shape, starting with fast decrease which then becomes slower.  Later, decreasing blocks show linear decrease, with flatter slope in the morning.
\item  The first decreasing blocks are the deepest: their endpoints have a value around 0.2. In the course of the night, these minimum values will increase to 0.25 or even to 0.3.
\item  Among constant blocks, wakeful activity has the highest level. It can vary between 0.25 and 0.6, depending on channel and person. 
\item  REM blocks are on a level slightly above 0.3 which remains unchanged or increases only slightly during the night. But a first short REM block can have a smaller level.
\item Constant S2 blocks appear between decreasing blocks and between a decreasing and a REM block. Their activity level is between the level of their neighbor blocks.
\item The level of the frontal channel exceedes the level of the parietal channel in wakeful and REM states. It is smaller at the end of decreasing blocks and on constant S2 blocks. 
\end{enumerate}

Now we are going to check these rules with several healthy and sick subjects, and note all deviations and individual differences.

\subsection*{Healthy probands}
Already in Figure \ref{fig:6} we see slight deviations from our rules. The 28 year old healthy woman n11 has low activity in the parietal channel, smaller than 0.3 already when she was awake.  Frontal activity is never smaller than parietal activity, only equal in slow wave sleep. 
Moreover, there is a gradual increase of the frontal channel in a decreasing block starting at 5:10 while the parietal channel stays below. n11 also shows many artifacts, and apparent arousals cause concatenation of decreasing blocks around midnight. Otherwise, it is similar to n5 with 8 hours of sleep without getting awake.

\begin{figure}[h!]
\begin{center}
\includegraphics[width=18cm]{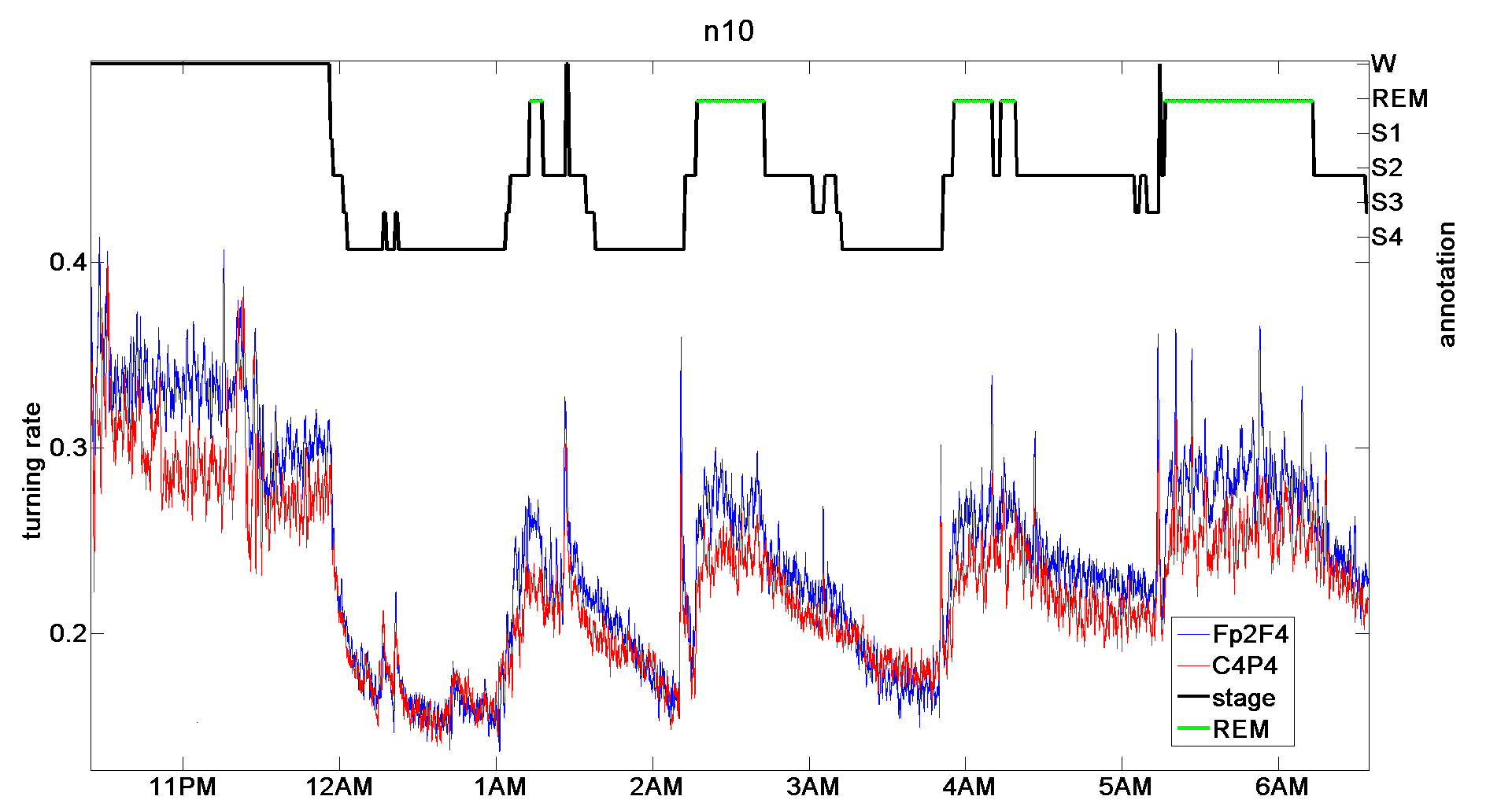} 
\end{center}
\caption{Continuous hypnogram of n10. Throughout, there is low activity level in both channels.}\label{fig:7}
\end{figure}

Figure \ref{fig:7} shows the 23 year old healthy man n10 which has an exceptional low level of activity throughout. Even when he was awake, activity was far below 0.4 and went down to 0.3 before the onset of sleep. Consequently, the REM level is only around 0.25. Thus the activity levels given in the rules above can hold only for most probands, not for all. There is a similar situation for amplitudes of EEG channels, which are very low for some healthy subjects \cite{ZH}, as well as for blood pressure and heart rate.  

\begin{figure}[h!]
\begin{center}
\includegraphics[width=18cm]{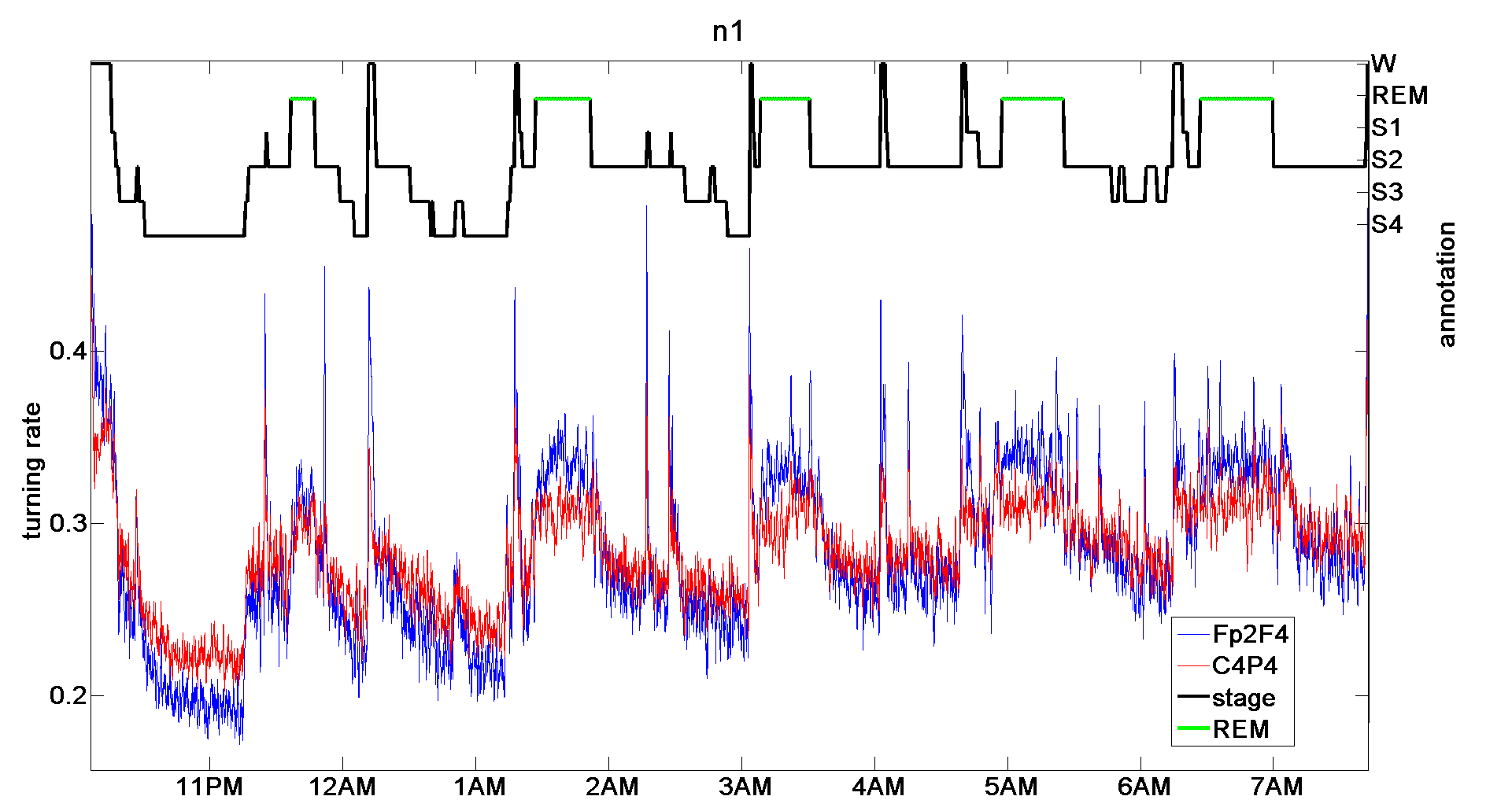} 
\end{center}
\caption{Continuous hypnogram of n1, in good agreement with all rules stated in the text.}\label{fig:8}
\end{figure}

The 37 years old healthy woman n1 shown in Figure \ref{fig:8} completely confirms all our rules although she briefly woke up six times during the night. In particular, constant S2 blocks can be identified before each REM phase which was not so clear for n10.

\begin{figure}[h!]
\begin{center}
\includegraphics[width=18cm]{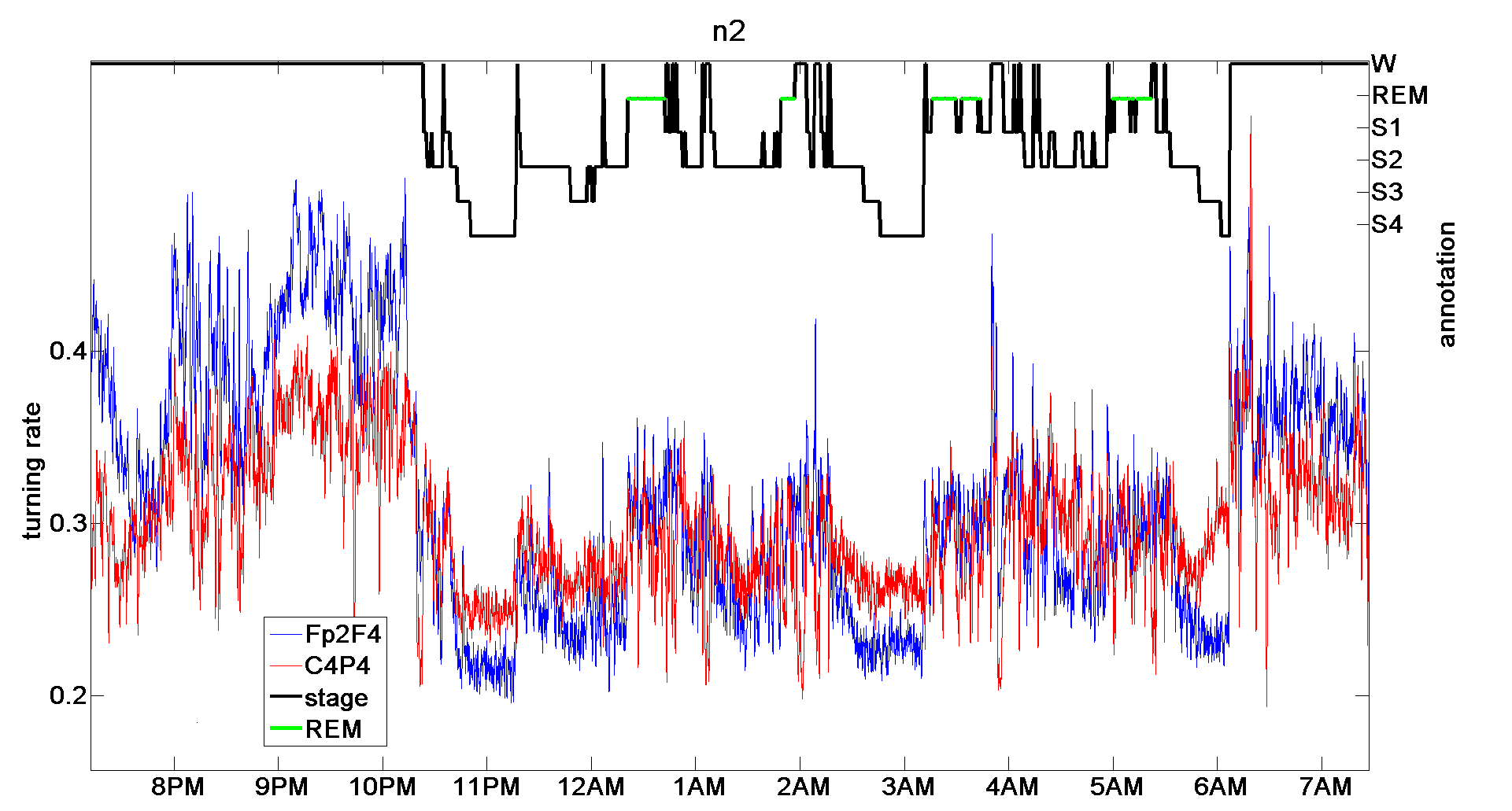} 
\end{center}
\caption{Continuous hypnogram of n2, with high parietal activity throughout sleep.}\label{fig:9}
\end{figure}

Finally, we consider the healthy controls n2 (34 years, male) and n3 (35 years, female) who did not sleep so well in the lab. 
The record of n2 in Figure \ref{fig:9} comprises 12 hours and allows us to observe the wakeful state. Vigilant activity in the morning is smaller than in the evening - this seems to be a general fact. The range of wakeful activity was between 0.3 and 0.5 in the frontal channel, and between 0.25 and 0.4 in the parietal channel. Had n2 a little nap before  8 pm? During sleep, the parietal channel did not come to rest, its value always being above 0.25. The frontal channel showed smaller values, but did approach 0.2 only once after 11 pm. The conventional hypnogram shows only three rather short S4 phases, and the values of the continuous hypnogram also tend to the assertion that the proband did not relax much although he was not  awake for longer time.
In the decreasing block around 6 pm, the parietal channel started to increase while the frontal one remained low, just opposite to Figure \ref{fig:6}.

n3 had strong frontal activity in the evening, with values around 0.5. Before sleep, this reduced to 0.4. The times when she was awake, around midnight and before 5 am, can be recognized from the large frontal values. During S4 blocks, frontal activity came to rest, and several times reached values below 0.2. In REM phases, however, frontal activity was not larger than parietal one, in contrast to all other probands. They were just equal. Several constant S2 phases can be identified, for instance at 1 am and before 3 am.
After 7 am, still annotated as S4, the frontal channel increases its activity for several minutes by a value of almost 0.2. 

\begin{figure}[h!]
\begin{center}
\includegraphics[width=18cm]{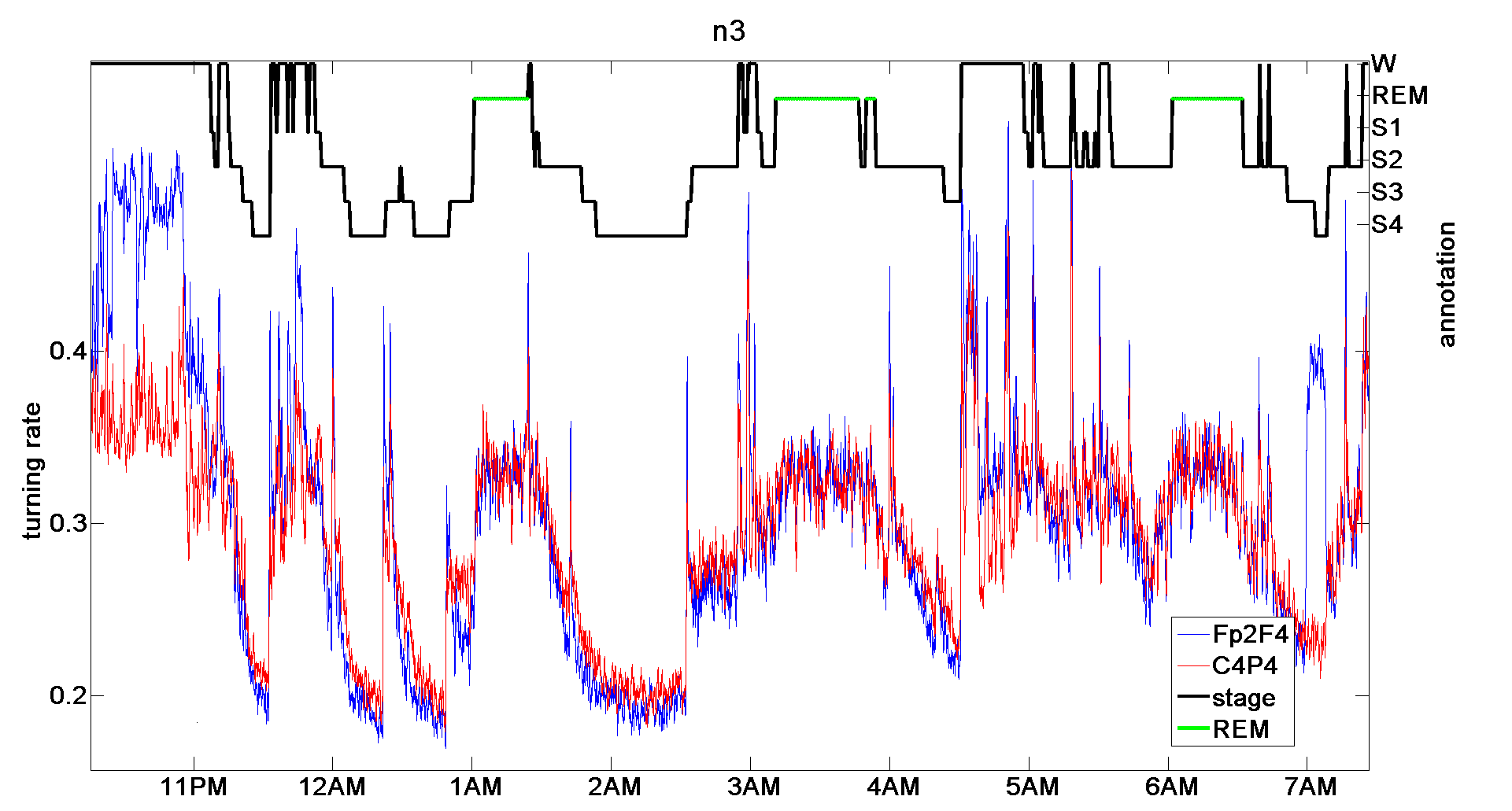} 
\end{center}
\caption{Continuous hypnogram of n3. Equal activity of both channels during REM. }\label{fig:10}
\end{figure}

We have seen a lot of individual differences. At the same time, our hypotheses have been confirmed: as rules, not as laws. What is most important: considering turning rate as brain activity, we obtained plausible observations and did not run into contradictions.

\subsection*{Sick subjects}
\begin{figure}[h!]
\begin{center}
\includegraphics[width=18cm]{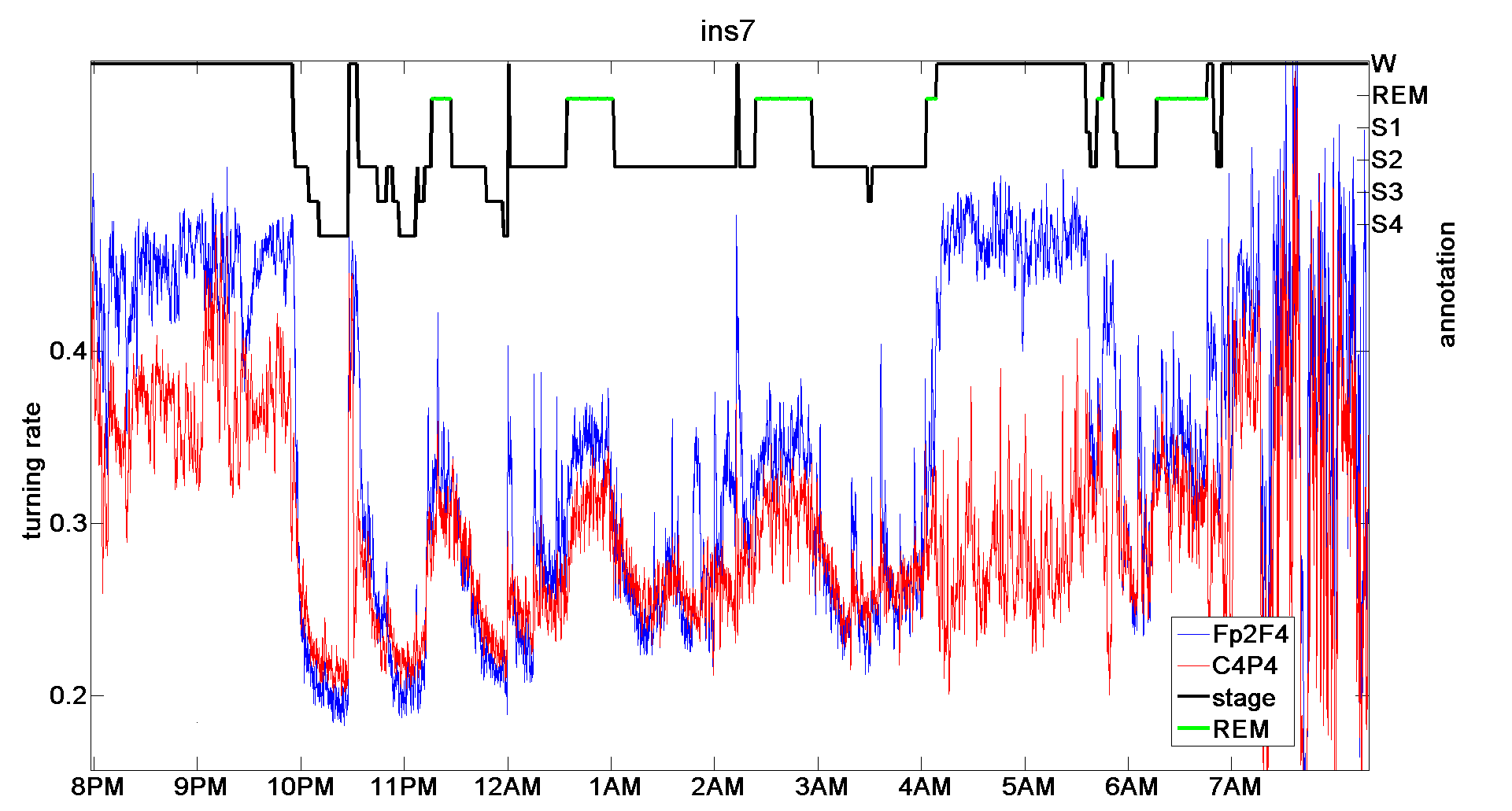} 
\end{center}
\caption{Continuous hypnogram of ins7. Exceptional frontal activity when awake in the early morning.}\label{fig:11}
\end{figure}

Within this paper we can give only a brief impression of the diagnostic potential of turning rates. We do not know the patients and their history, and should avoid speculation. A first example is Figure \ref{fig:11}, a woman of 47 years  suffering from insomnia. The continuous hypnogram up to 4 am shows normal sleep, in complete agreement with our rules and the healthy controls. But from 4:10 up to 5:40 the person is awake, and the frontal activity in this time is higher than before sleep and after sleep. The sleep architecture is normal but the person seems to have problems to think about when she is awake at night.

\begin{figure}[h!]
\begin{center}
\includegraphics[width=18cm]{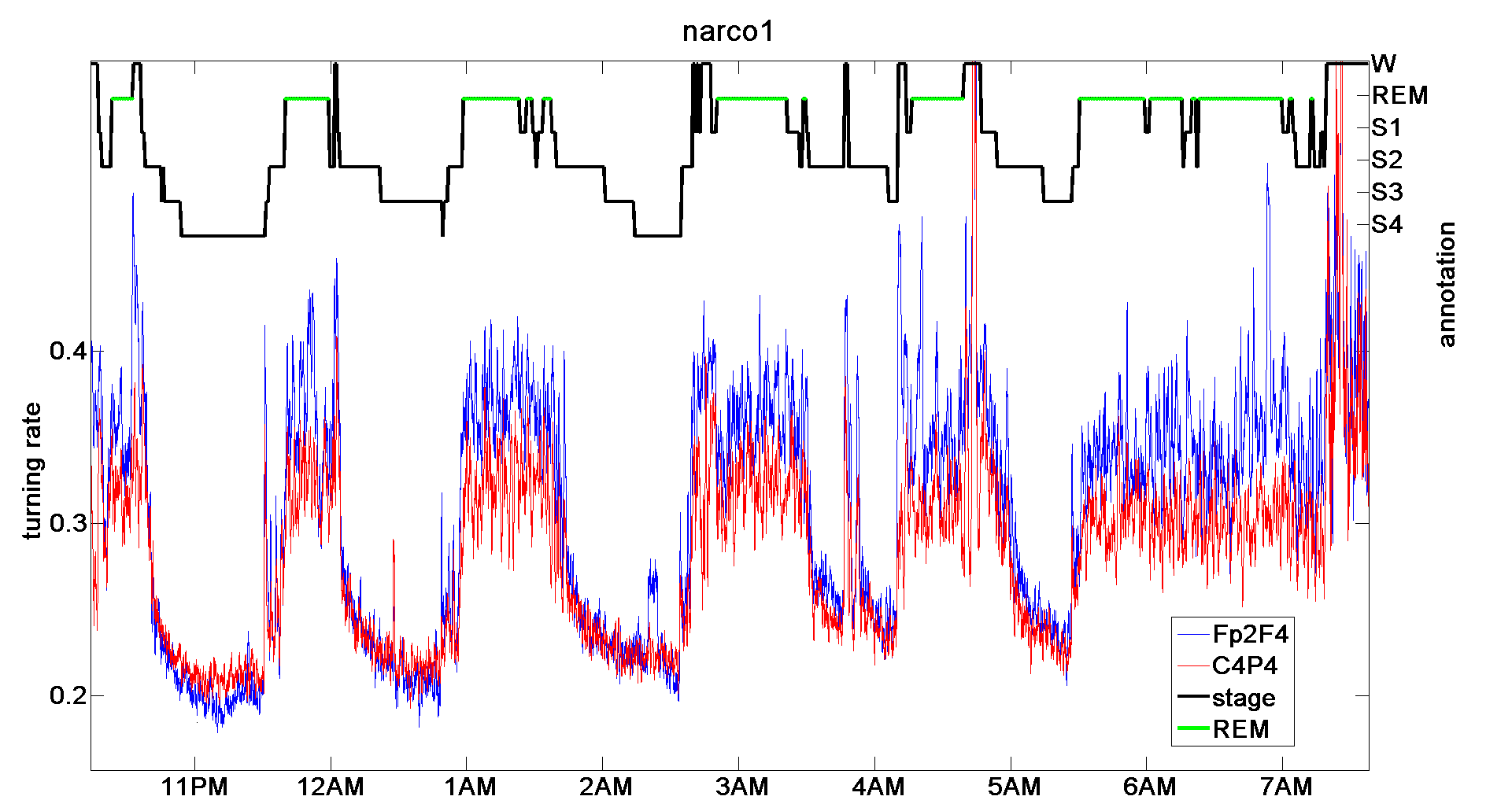} 
\end{center}
\caption{Continuous hypnogram of narco1; Very high REM activity with large variation.}\label{fig:12}
\end{figure}

Three other examples from the narcolepsy group will be given. Patient narco1, a 29 year old woman, has a continuous hypnogram which fulfils all our rules in an ideal way (Figure \ref{fig:12}). We have six REM phases with frontal activity larger than parietal one. The activity level of the REM phases, as well as the variation during a REM phase, are larger than for all our healthy subjects. There are short constant S2 blocks before each REM phase. All decreasing blocks have parabolic shape: the patient gets into deep sleep very fast, and three decreasing blocks approach the level 0.2. There is a little excursion of the frontal channel at 2:30 am, and one can ask why we get so immediately into the first REM phase at 10:30 pm. But we cannot say there is an abnormal sleep architecture. We could only conclude: this person perhaps sleeps too well.

\begin{figure}[h!]
\begin{center}
\includegraphics[width=18cm]{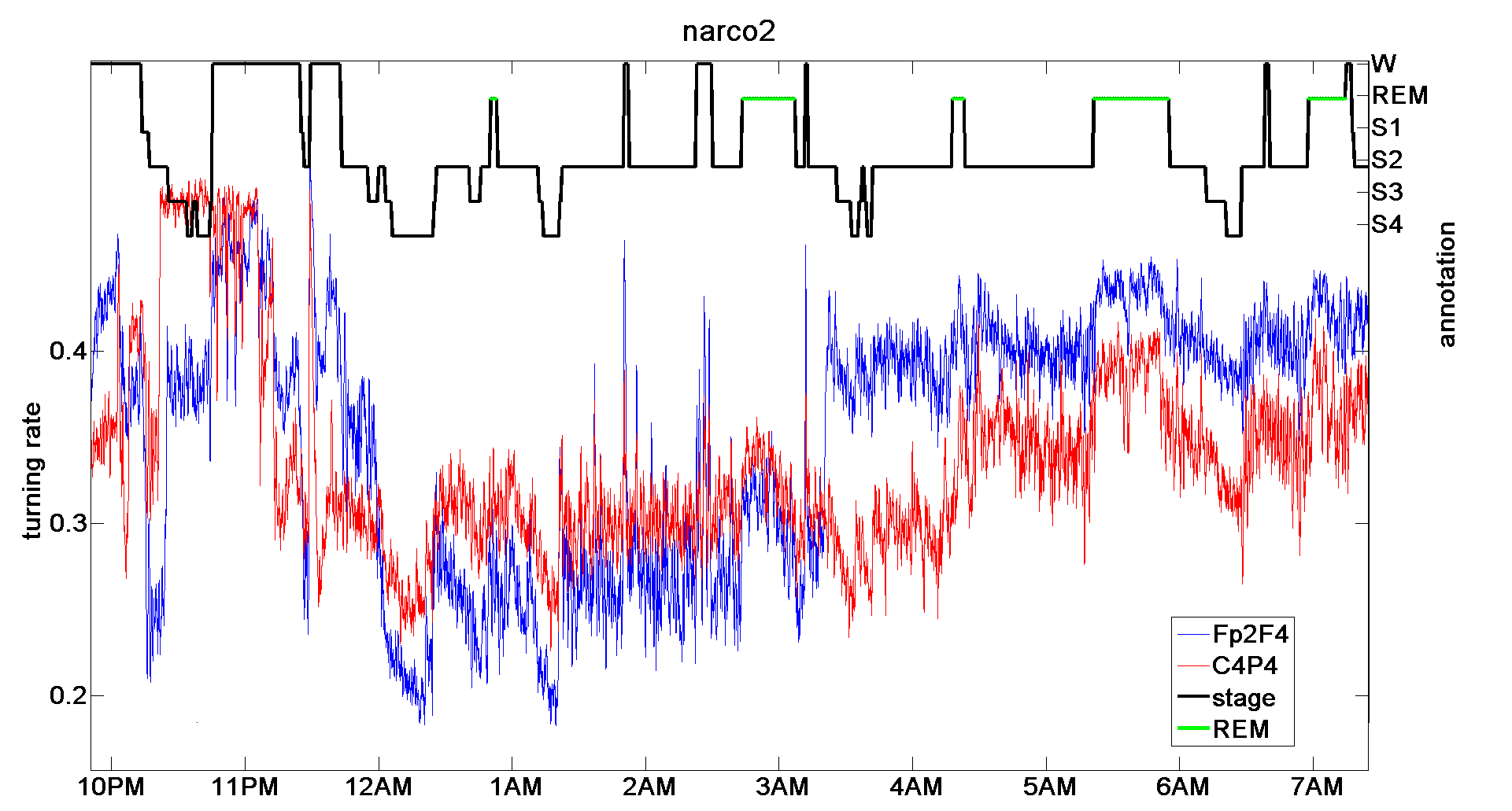} 
\end{center}
\caption{Continuous hypnogram of narco2. Degenerate sleep after 3:20 am.}\label{fig:13}
\end{figure}

This will be contrasted with patient narco2, a woman of 44 years shown in Figure \ref{fig:13}. The conventional hypnogram contains not much slow wave sleep, and there is an isolated sleep phase between 10 and 11 pm. Otherwise it does not look worse than for the healthy subjects in Figures \ref{fig:9} and \ref{fig:10}.  Moreover, for the time between midnight and 3:20 am, continuous and conventional hypnograms coincide in a similar way as for healthy subjects, with the exception that the frontal channel shows smaller activity within REM phases. The REM phases annotated 4:20, 5:30 and 7 am are reflected in the continuous hypnogram by higher turning rates.

However, from 3:20 am to the end of the record, the structure of the continuous hypnogram is totally different from the time before. The turning rates in the frontal channel are mostly above 0.4, the expected level of wakeful acitivity. A similar remark holds for the parietal channel after  4:30 pm. The whole period after 3:20 am should be classified as a state between REM and wakefulness, not a sleep phase. There is a serious disturbance of sleep architecture. Nevertheless, we can still recognize constant and slightly decreasing blocks, with activity in both channels running in a parallel way, as in the wakeful state in Figures \ref{fig:7}, \ref{fig:9}, and \ref{fig:11}.  When we consider the continuous hypnogram of narco2 between 10 pm and midnight, we see a lot of jumps of activity in both channels, in particular an exceptionally strange level of 0.5 in the parietal channel between 10:25 and 11 pm. We would never classify this time as deep sleep. This continuous hypnogram definitely shows degenerate sleep.

\begin{figure}[h!]
\begin{center}
\includegraphics[width=18cm]{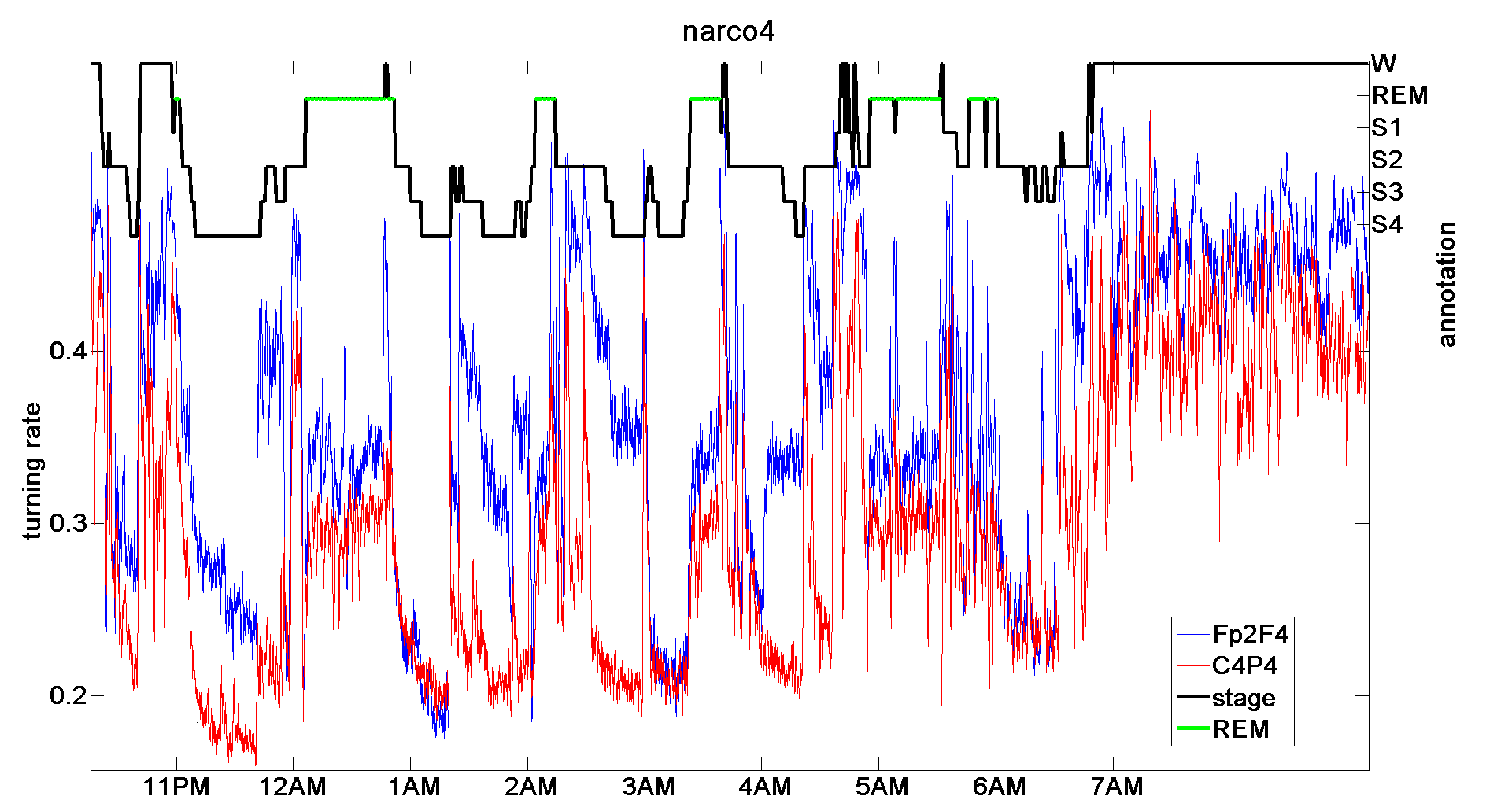} 
\end{center}
\caption{Continuous hypnogram of narco4. Jumps to high frontal activity during slow wave sleep.}\label{fig:14}
\end{figure}

Figure \ref{fig:14} shows a similar case. The conventional hypnogram narco4 (43 years, male) looks similar as for healthy controls. It coincides with the continuous hypnogram on REM phases, which all show the same level of turning rates, with dominating frontal channel. When we restrict ourselves to the parietal channel, we could say that continuous and conventional hypnograms agree as in healthy cases. However, the frontal channel behaves in a very erratic manner: it jumps to levels of REM or even wakeful activity when we still have a decreasing phase of slow wave sleep - for instance at midnight, between 1 and 2 am, very clearly between 2 and 3 am, and again around 4:30 am. This patient also seems to have disordered EEG sleep activity, and one could go ahead by studying more EEG channels with turning rates.

\subsection*{Remarks on $\tau $ waves}
For our first healthy case n5, we studied the continuous hypnogram in detail and found $\tau $ waves with a wavelength between 30 s and 2 min which where most obvious at the onset of sleep and in constant S2 blocks. These waves appear in both channels in an almost parallel way so that we can conclude that they have a source in central brain. For n5, these waves were suppressed in REM phases and slow wave sleep. What about the other subjects? 

The phenomenon of $\tau $ waves seems universal. We have found them in all healthy and sick persons which we studied. They are always visible at the onset of sleep and at the beginning of most decreasing blocks which lead to deeper sleep, and at these places they can be very large. 
There are also constant S2 blocks in almost all of our hypnograms. Some of them are rather short. In longer constant S2 blocks we regularly find $\tau $ waves with small amplitude and strong synchronization between the two channels. In all cases, the morphology of $\tau $ waves resembles $\theta$ or $\delta$ waves rather than the more sinusoidal $\alpha$ waves. The shape of these waves is of course influenced by our choice of the moving average filter.

Within slow wave sleep, the turning rate does not oscillate much, and $\tau $ waves are suppressed. In REM sleep, as well as in wakeful activity, some of our subjects showed clearly synchronized $\tau $ waves. On the whole, synchronized $\tau $ waves seem to represent the rule rather than an exception.

\begin{figure}[h!]
\begin{center}
\includegraphics[width=18cm]{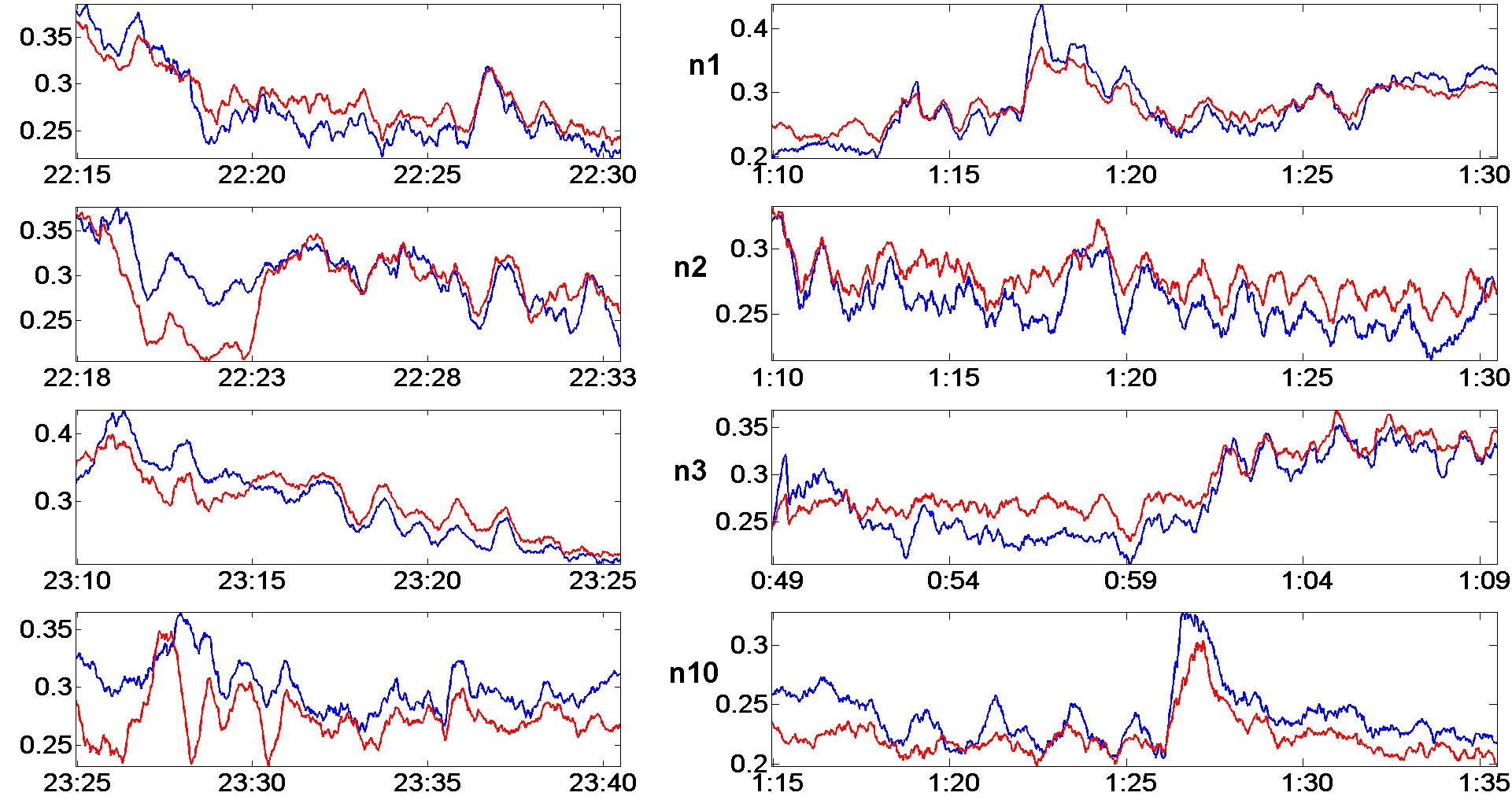} 
\end{center}
\caption{Details of the continuous hypnogram of healthy probands. On the left, the onset of sleep is shown for 15 minutes, on the right a constant S2 block within a window of 20 minutes. In all cases, there are synchronous $\tau $ waves in both channels.  }\label{fig:15}
\end{figure}

There is no place here to study $\tau $ waves in greater detail. This should be done with a comprehensive set of EEG channels, in order to understand in which regions of cortex $\tau $ waves are particularly strong. Here we just add Figure \ref{fig:15} which lists the onset of sleep and a constant S2 block around 1 am for four more healthy subjects. The synchronization of our two channels in all these cases is a strong argument for the subcortical origin of $\tau $ waves.

\section*{Discussion}
This paper introduces a new way to evaluate EEG measurements. It is claimed that the count of turning points in 1s epochs provides a good proxy for activity in the corresponding region of brain. Certainly, this concerns only a very special aspect of cortex activity. However, in most clinical applications of EEG, like epilepsy, stroke, brain injury, psychic disorders etc., and sleep, {\it all} aspects of brain activity are affected. Thus if our claim is true, turning rate has a huge potential for medical applications. The method can easily be implemented in real time so that the activity in the last epoch can be observed during measurement.

The count of turning points admits a considerable statistical error. Smoothing over some time, typically 30 seconds, is necessary, and only relatively slow changes of activity can be studied. For this reason, sleep EEGs of at least six hours duration were selected to verify the claim by example. The activity curve of an EEG channel was called continuous hypnogram. It resembles the hypnogram of annotations of sleep stages by expert physicians in an impressive way.

The coincidence of the two types of hypnograms is remarkable for two reasons. On the one hand, sleep stage annotation is an art which is based on a complicated set of rules and requires considerable experience and training. In contrast, the continuous hypnogram is based on a simple calculation which involves no human or machine intelligence,  no special preprocessing, and in this paper no artifact  treatment.  On the other hand, sleep stage annotation bases on visible graphic elements of the EEG, like $\alpha$ and $\delta$ waves, spindles, K-complexes etc. Turning rate is based on comparison of measurements with a distance of less than 10 ms, which in clinical practice is considered as
`background noise' of the EEG. Although both methods exploit different features, they lead to similar estimates of sleep depth.

Previous studies on permutation entropy \cite{KL,NG,Ba17a,Bru,Fe,Mor,OSD} have shown that EEG background noise contains important information. Their methodology was simplified and improved here. It can be proved with mathematical rigor that turning rate is a main component of permutation entropy \cite{Ba15}. There are other components of smaller importance which have been neglected for the present paper. The main conclusion of these studies is that there is important content in the fine structure (`background noise') of the EEG. Thus it makes sense to measure with high sampling frequency and avoid excessive filtering which can destroy the fine structure. This argument is confirmed by the present study.

All data of the paper were taken from the CAP database of Terzano et al. \cite{Te} at physionet \cite{physio}. Six healthy controls, one insomnia and three narcolepsy patients were studied and the continuous hypnograms presented here. These rather promising ten examples give an impression of the variety in the healthy case and the diagnostic potential  in the narcolepsy case.  Nevertheless, the number of cases is too small to obtain a definite judgement on the turning rate method. Further studies are required. 

This paper used a particularly simple setting with only two bipolar EEG channels Fp2F4 and C4P4.
Different EEG channels and montages have to be investigated in order to check spatial effects. Moreover, there are different options for calculating turning rates. Here we took the distance $d=4$ for data with sampling frequency 512 Hz, with the purpose of presenting a most simple and transparent method. 
In \cite{Ba17a} we took the average over $d=2,3,...,20.$ Now we recommend to avoid $d>10,$ as a result of numerical experiments. Nevertheless, an average, or better a weighted average over several $d$ could be more efficient.

Turning rates are estimated statistically from non-overlapping or overlapping epochs, and their variance creates an unavoidable statistical error. In this study, 1s epochs were used and then smoothed out over 30s. At present stage this is sufficient for sleep medicine. For other applications, like evoked and event-related potentials, it would be desirable to detect fast changes of brain activity. Averaging is a traditional way to reduce variance. Besides averaging over delays, one can average over more time points when sampling frequency is increased, or over several EEG channels. The effect depends on correlations of the terms in the mean, and has to be checked by experiment.

The statistical nature of turning rates allows us to study very slow waves, since the new time series of epochs is much shorter than the original series of values. Oscillations with wavelength between 30s and 2 min were found in all ten subjects of the present study. They were termed $\tau $ waves for short and seem to be always present within sleep stages S1 and S2. Their synchronized appearance in both channels is an argument for subcortical origin. 

In the literature, waves with a length of at least 10s are called infra-slow. The term was introduced in 1957 by Aladjalova  \cite{Aladjalova57} who found oscillations of wavelength 7-10s and 30-90s in steady potential differences of implanted electrodes in rabbits. Many electrophysiological animal studies have confirmed the existence of infra-slow waves, in particular at the thalamus, and slices of isolated thalamic relay nuclei show spontaneous infra-slow waves in vitro \cite{Hughes11}, In recent years, a lot of fMRI studies have verified infra-slow waves in the intensity of the BOLD (blood-oxygen-level dependent) signal, in many regions of human brain at rest \cite{Zuo10,Picchioni11,Mitra15}.

Infra-slow oscillations were also observed in intracranial records (ECoG) and local field potentials  \cite{He08,Nir08}. 
In ordinary EEG records they appear as periodic occurence of spindles and CAP sequences \cite{Te,Parrino06}, 
and as modulation of the power of traditional frequency bands \cite{Mantini07}. It seems curious that turning rates with a delay of 8ms, corresponding to frequencies beyond the $\gamma$ band, will reveal infra-slow waves. However, this is in accordance with the literature which states strong correlation between infra-slow waves and power of $\gamma$ frequencies  \cite{Vanhatalo04,Mantini07,He08,Monto08}.
The full-band EEG recordings obtained with sophisticated equipment \cite{Vanhatalo05} by Vantahalo et al.  \cite{Vanhatalo04} and Monto et al. \cite{Monto08} are most similar to our $\tau $ waves since they also present the waves themselves, not their Fourier spectrum.  These authors found infra-slow waves in the original EEG series while we observe them in the turning rate.

According to the terminology of Penttonen and Buzsaki, wavelength between 30 and 120s is denoted slow-5 band \cite{Buzsaki04,Zuo10} while most of the above papers study the slow-4 band with wavelength between 10 and 30s. Due to our moving average over 30 epochs of 1s, we could not observe slow-4 waves. However, $\tau $ waves in continuous hypnograms seem a most simple way to observe slow-5 oscillations. We did not intend to study such waves. They just came as a byproduct of the continuous hypnogram. It is not clear how they relate to the diverse oscillations quoted above. We raised more questions than answers.

The paper suggests that turning rates and continuous hypnograms can become a standard tool, a biomarker \cite{Lew, Pen} for EEG evaluation and sleep medicine, similar to blood pressure and laboratory measurements. Actually, this could save physicians a lot of routine work, and due to the ever-increasing amount of long-term biodata, computer tools become inevitable. But even if the claims of this paper should turn out to be true, it is still a long way to an acceptable biomarker. A first step is to check whether turning rates are reproducable in repeated measurements under similar conditions. The variation for healthy subjects should not be too large, and deviations of sick subjects from the healthy state should be well detectable. Dependence of turning rates on the prefiltering options must be determined and, if possible, understood theoretically. Different options of turning rates and related parameters have to be compared. An optimized and standardized version of `brain activity' has to be defined which convinces the community of EEG physicians. Can the suggested method stand these tests? It is worth a trial. For screening long-term EEG, it seems helpful already in its present form.

\subsection*{Remarks and Acknowledgments}
The author declares that all work for this paper was done by himself, and that there is no potential conflict of interest. 
He gratefully acknowledges the use of the CAP sleep database of Terzano et al. \cite{Te} accessible at physionet \cite{physio}. The author thanks Bernd Pompe for fruitful discussions, critical advice, and encouragement.


\bibliographystyle{frontiersinHLTH&FPHY} 
\bibliography{time1}

\end{document}